\documentclass{IEEEtran}
\usepackage{cite}
\usepackage{amsmath,amssymb,amsfonts}
\usepackage{algorithmic}
\usepackage{graphicx}
\usepackage{textcomp}
\usepackage{tikz}
\usepackage{pgfplots}
\pgfplotsset{compat=1.18}
\usetikzlibrary{arrows.meta, decorations.pathreplacing}
\usepackage[caption=false,font=footnotesize]{subfig}
\begin{document}
\title{Analytical Framework of Radial Resolution for Near-Field Communications}
\author{{\IEEEauthorblockN{Viviana Centritto, Sotiris Droulias and Angeliki Alexiou,~\IEEEmembership{Member,~IEEE}} }
\thanks{The authors are with the Department of Digital Systems, University of Piraeus, Piraeus 18534, Greece (Corresponding author: Viviana Centritto, \mbox{e-mail:vcentritto@unipi.gr).}}}

\maketitle

\begin{abstract}
As extremely large antenna arrays (ELAAs) become central to next-generation wireless systems, the transition into the near-field propagation regime enables the exploitation of spherical wavefronts for radial-domain beamfocusing. This capability is pivotal for emerging applications requiring precise spatial isolation, such as Space Division Multiple Access (SDMA), hierarchical localization and advanced sensing. However, fully realizing these technologies requires specific design rules to dimension multi-user systems without relying on computational expensive full-wave simulations. To address the gap in modeling contiguous focal regions with controllable radial resolution, this paper expands the Angular Spectrum Representation (ASR) approach to propose a comprehensive analytical framework. Through the introduction of a tunable inter-beam overlap parameter $\rho$, we derive closed-form expressions to synthesize multiple focal regions, providing the flexibility to tailor their radial resolution. Furthermore, the resolution capabilities are analyzed to characterize the interplay between key operational variables, such as the transmitter size, beam radius and operation frequency. System-level assessment of per-user and sum-rate spectral efficiencies across varying signal-to-noise (SNR) regimes reveals how the inter-beam overlap dictates a fundamental trade-off between user capacity and inter-user interference, delivering design guidelines for future near-field communications.
\end{abstract}


\section{Introduction}
\label{sec:introduction}
The evolution of next-generation wireless systems toward extremely large antenna arrays reshapes the boundary between the near-field and far-field propagation regimes \cite{bjornson2019massive, zhang20236g, ye2024extremely}. As the physical aperture of the array increases, this boundary extends significantly. Consequently, end-users, sensor nodes and environmental scatterers traditionally assumed to be in the far-field are increasingly located within the near-field of the array, allowing the exploitation of new spatial properties \cite{liu2024near ,liu2025near}. In this operating regime, the plane-wave approximation is no longer applicable and the full spherical wavefront must be taken into account. This enables phase variations across the array aperture that depend on the angle and radial distance to the observation point, rather than being dictated by the angular direction alone \cite{zhou2015spherical ,ouyang2024primer}. 

The joint angle-distance dependence in the near-field regime enables a critical capability: beamfocusing, also known as range-dependent or finite-depth beamforming \cite{bjornson2021primer, zhang2022beam, kosasih2024finite}. The array concentrates its radiated power into a localized, three-dimensional focal spot at a prescribed distance and direction, instead of steering a beam of constant angular width that propagates with monotonically decreasing amplitude, as occurs in conventional far-field beamforming. Therefore, beamfocusing introduces an additional spatial degree of freedom that can be exploited. For example, users sharing the same angular direction can be served simultaneously through range-domain discrimination, in a near-field multiple access scheme \cite{decarli2021communication, zuo2023near, wu2022multiple, cui2022near, ouyang2024primer, droulias2025orthogonal, droulias2025orthogonalcodebook}. Because the array can synthesize multiple controllable focal regions (or non-overlapping spatial zones), these users can be distinguished and served with reduced inter-user interference.

Fully exploiting this capability requires a rigorous characterization of the radial resolution of these focal areas. The need to design systems tailored to advanced sensing, localization and range-domain multiple access \cite{cong2024near, gavriilidis2025near, mozaffarikhosravi2025localization, monemi2025toward, yang2025beam, li2026robust}  has made defining this resolution a central research focus in recent years. Among the early works on beamfocusing for wireless connectivity \cite{yurduseven2020intelligent, zhang2021beam, zhang2022beam, droulias2022reconfigurable, ding2023resolution}, an analytical framework for assessing the radial resolution of the focal areas was first reported in \cite{droulias2022reconfigurable} and further analyzed in \cite{stratidakis2023beam, droulias2024near}. In \cite{droulias2022reconfigurable, stratidakis2023beam} the authors applied the Angular Spectrum Representation (ASR) approach to Reconfigurable Intelligent Surfaces (RISs), by treating the RIS as a continuous surface. This approach yielded analytical expressions for the modeling of beam steering, optical RIS placement and beamfocusing. Expanding on these principles, \cite{droulias2024near} proposed an ASR-based analytical model to assess beamfocusing at oblique incidence and reflection angles. The work derived a closed-form expression for the spatial power density, alongside key performance metrics such as the Full-Width-Half-Maximum (FWHM), which quantifies the longitudinal spatial extent of the focal region. Leveraging on these findings, in \cite{stratidakis2025near} the same authors demonstrated a hierarchical localization algorithm for estimating the location of a user. The algorithm uses beamforming and beamfocusing, to estimate the angle and distance of the user with controllable angular and radial resolution, respectively.

An alternative route, based on the Fresnel approximation of the Huygens-Fresnel integral was later followed in \cite{kosasih2024finite}, in which the authors provided an analytical model for evaluating the 3 dB beam depth (BD), deriving closed-form array gain and BD expressions across rectangular and circular array geometries. Their study demonstrates how aperture length, array shape, and steering angle affect the range-domain resolution, proving that nonoverlapping 3~dB BD focal spots enable co-angular users to be multiplexed in the range-domain with reduced multiuser interference. Building upon these principles, the authors in \cite{wachowiak2025sizing} advanced these findings by proposing closed-form solutions adapted to various array geometries to quantify key near-field performance metrics and their scaling behavior as a function to the array aperture. In particular, the authors established that the minimum achievable BD asymptotically saturates as the aperture increases, calculated the NF region span, and estimated the maximum number of 3~dB resolvable beamspots that can be accommodated within this extent. Furthermore, the authors in \cite{manohar2024resolution}  evaluated the scalar Fresnel- Kirchoff diffraction integral to derive analytical solutions for both transverse and longitudinal resolutions. Their analysis provides insights into how 3D focal spot dimensions scale with array geometry, focal distance, and operating frequency. 

While previous studies have established foundational models for characterizing radial resolution capabilities, they predominantly address isolated focal regions. Consequently, a critical research gap remains: the lack of an analytical model that synthesizes contiguous focal areas, tailored to serve multiple users in advanced near-field scenarios. In this work, we address this gap by extending the ASR approach as an electromagnetic modeling tool. By capturing the full-wave propagation dynamics of the near-field, we derive a recursive analytical framework for the arrangement of multiple controllable focal regions along the radial domain. Beyond this primary capability, the explicit mathematical formulation of the model reveals how design variables interact to govern the spatial radial resolution. These insights can be translated into design guidelines for dimensioning near-field beamfocusing systems, avoiding the need for computationally expensive full-wave simulations or numerical sweeps. The main contributions of this paper are summarized as follows: 

\begin{itemize}
    \item We design an analytical model to arrange $N$ contiguous focal regions, enabling the simultaneous accommodation of $N$ users along the radial domain. To achieve precise spatial isolation, this framework introduces the tunable design parameter $\rho$ that dictates the radial separation and inter-beam overlap.
    \item We derive closed-form expressions to maintain a uniform radial resolution across all focal areas. Furthermore, the model is generalized to accommodate heterogeneous scenarios, providing the capability to independently tailor the resolution of each focal area.
    \item We analytically establish the spatial boundaries of the system by formulating the optimal focal distances for the outermost beam, constrained by user capacity or finite aperture limits, and the minimum focal distance necessary to ensure valid paraxial approximation propagation.    
    \item We conduct a system-level evaluation to assess the performance of the $N$ synthesized focal areas, quantifying both the per-user and sum-rate spectral efficiencies. In addition, an asymptotic analysis of the low- and high-SNR regimes demonstrates how the inter-beam overlap parameter $\rho$ dictates a fundamental trade-off between the number of simultaneous users supported and the inter-user interference.
\end{itemize}

\section{System model and analytical framework}
\label{sec:framework}
\begin{figure}
    \centering
    \includegraphics[width=0.95\linewidth]{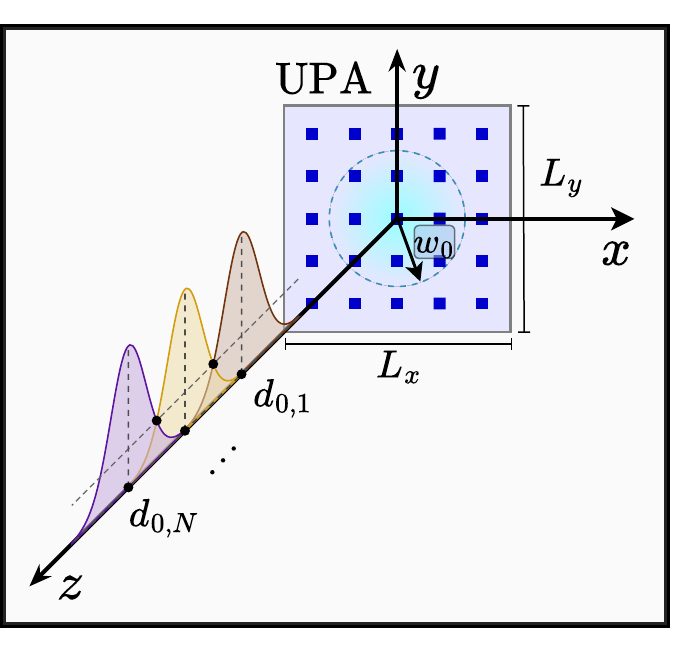}
    \caption{UPA with $L_{x}\times L_{y}$ elements operating in the near-field regime. The array, characterized by a beam radius $w_0$, synthesizes $N$ contiguous focal regions along the radial propagation direction ($z$-axis) centered at focal distances $d_{0,1}, \dots, d_{0,N}$.}
    \label{fig:system_model}
\end{figure}

Consider a transmitter equipped with a uniform planar array (UPA) centered at the origin of the coordinate system and located on the $xy-$ plane. As illustrated in Fig. \ref{fig:system_model}, the UPA comprises a total of $L$ radiating elements, with $L_x$ and $L_y$ distributed along the $x-$ and $y-$ axis, respectively. The electrically large aperture of the UPA enables near-field spherical wavefront manipulation to focus the radiated beam at a focal distance $d_0$. Consequently, a spatially confined focal region is generated, which can be allocated to a specific user.
The field profile at the input plane ($z=0$) is formulated as a continuous function comprising amplitude and phase components
\begin{equation}\label{eq:field_UPA}
    E^{\text{UPA}}(x,y) = A(x,y) \, e^{j \Phi(x,y)}.  
\end{equation}
The amplitude $A(x,y)$ is modeled as a Gaussian distribution characterized by the beam radius $w_0$ and the electric field magnitude $E_0$, as expressed in (\ref{eq:A(x,y)}). To focus the beam along the propagation direction ($z$-axis) with no beam-steering, the phase profile $\Phi(x,y)$ imposed by the array requires a spherical curvature. For most cases of practical interest \cite{droulias2024near}, this phase can be approximated as shown in (\ref{eq:Phi(x,y)}), where $k=\frac{2\pi f}{c}$ is the free-space wavenumber at the operating frequency $f$
\begin{equation}\label{eq:A(x,y)}
    A(x,y) = E_0 \, e^{- \frac{x^2+y^2}{w_0^2}},  
\end{equation}
\begin{equation}\label{eq:Phi(x,y)}
    \Phi(x,y) \approx -k \left( d_0 + \frac{x^2 + y^2}{2 d_0} \right).
\end{equation}

To characterize the beam's propagation in the near-field region, we employ the ASR approach. This technique decomposes the source field at the aperture plane ($z = 0$) into superposition of plane waves, whose two-dimensional Fourier spectrum determines the field distribution at any observation plane $z>0$ \cite{droulias2022reconfigurable, droulias2024near}. First, the spatial frequency representation of the UPA is obtained via the Fourier transform of the aperture field
\begin{equation}\label{eq:F1_UPA}
    \hat{E}(k_x,k_y)=\frac{1}{4\pi^2}\iint_{-\infty}^{+\infty}E^{\text{UPA}}(x,y)e^{-j(k_xx+k_yy)}dxdy,
\end{equation}
where $k_x$ and $k_y$ represent the transverse spatial frequencies. 

To ensure consistency, the radiated field $E(x,y,z)$ must satisfy Maxwell's equations, which for free-space reduce to the  Helmholtz equation $(\nabla^2+k^2)E(x,y,z)=0$. By expressing $E(x,y,z)$ in terms of its Fourier representation and substituting it into the Helmholtz equation, we determine its evolution along the propagation axis $z$
\begin{equation}\label{eq:evolve}
    \hat{E}(k_x,k_y;z)=\hat{E}(k_x,k_y)e^{jk_zz},
\end{equation}
where the longitudinal spatial frequency $k_z$ is defined as
\begin{equation}\label{eq:k_z}
    k_z=\sqrt{k^2-k_x^2-k_y^2}.
\end{equation}

Applying the inverse Fourier transform to (\ref{eq:evolve}) yields the \textit{angular spectrum representation}, providing an exact expression for the field at any observation point $(x,y,z)$
\begin{equation}\label{eq:ASR}
   {E}(x,y,z)=\iint_{-\infty}^{+\infty}\hat{E}(k_x,k_y)e^{j(k_xx+k_yy+k_zz)}dk_xdk_y.
\end{equation}
An analytical solution was found in \cite{droulias2022reconfigurable} by solving (\ref{eq:ASR}) for beam focusing solely along the $z$-axis. Assuming the wave-vector $\mathbf{k} = (k_x,k_y, k_z)$ is almost parallel to the $z$-axis, its transverse components $k_x$ and $k_y$ are significantly smaller than the longitudinal component $k_z$. This condition allows a Taylor expansion of $k_z$ around the direction of propagation ($k_x= 0 $ and $k_y = 0$). This expansion results in the well-known paraxial approximation (\ref{eq:k_z})
\begin{equation}\label{eq:paraxial}
    k_z \approx k - (k_x^2 + k_y^2)/2k.
\end{equation}

By substituting the approximations from (\ref{eq:Phi(x,y)}) and (\ref{eq:paraxial}) into (\ref{eq:ASR}), we can find the square magnitude of the propagated field at any observation point $(x,y,z)$ as
\begin{equation}\label{eq:E(x,y,z)_magnitude_squared}
    |E(x,y,z)|^2 =  |E_0|^2 \frac{w_0^2}{w^2(z)}\exp \left [-2 \frac{x^2+y^2}{w^2(z)} \right],
\end{equation}
where $z_R = \frac{k w_0^2}{2}$ is the Rayleigh range associated with the array's beam radius $w_0$, and $w(z)$ is the $z$-dependent beam radius
\begin{equation}\label{w_z}
w(z) = w_0 \sqrt{\left(1-\frac{z}{d_0}\right)^2+ \left(\frac{z}{z_R}\right)^2.}
\end{equation}
Consequently, the spatial power density delivered by the UPA can be evaluated as $S=\vert E\vert^2/2\eta_0$, where $\eta_0$ is the free-space wave impedance. Evaluating along the longitudinal axis, the power density simplifies to
\begin{equation}\label{eq:power_density}
    S(z)=\frac{\vert E_0\vert^2}{2\eta_0}\frac{w_0^2}{w^2(z)}.
\end{equation}
By differentiating $S(z)$ with respect to $z$ and setting the result to zero, we can determine the position $z_{\text{peak}}$ at which the beam's intensity reaches its maximum value
\begin{equation}\label{eq:z_peak}
    z_{\text{peak}} = \frac{d_0 z_R^2}{z_R^2 + d_0^2} = \frac{d_0}{1 + (\frac{d_0}{z_R})^2},
\end{equation}
replacing this location back into (\ref{eq:power_density}) yields the maximum achievable power density $S_{\max}$
\begin{equation}\label{eq:s_z,max}
    S_{\max} = S(z = z_{\text{peak}}) = \frac{|E_0|^2}{2 \eta_0} \, \left[ 1 + \left(\frac{z_R}{d_0} \right)^2 \right].
\end{equation}

\begin{figure}[!t]
    \centering
    \subfloat[]{%
        \includegraphics[width=0.95\linewidth]{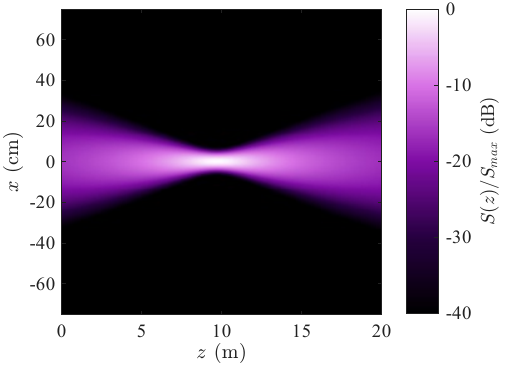}%
        \label{fig:spatial_distribution}
    }\\
    \vspace{0.1cm}
    \subfloat[]{%
        \includegraphics[width=0.95\linewidth]{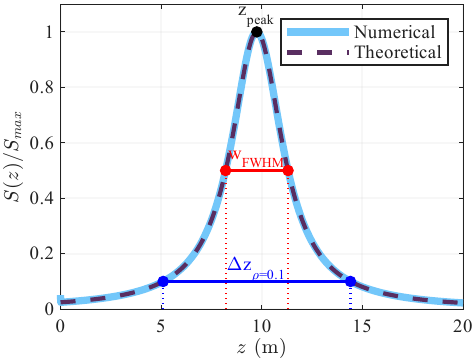}%
        \label{fig:cross-section}%
    }
    \caption{Normalized power density distribution of a single focused beam.         (a)~Cross-section rendered in the $xz$-plane. (b) Profile along the $z$-axis comparing the theoretical framework against numerical propagation. Key parameters such as the peak location $z_{\text{peak}}$, radial extent $w_{\text{FWHM}}$, and spatial interval $\Delta z_{\rho=0.1}$ are highlighted.}
    \label{fig:single-beam}
\end{figure}

To validate the analytical power density expression derived in (\ref{eq:power_density}), Fig. \ref{fig:single-beam} contrasts the proposed theoretical framework with a numerical model baseline. This numerical model serves as a reference by computing the exact discrete superposition of spherical waves radiated from each UPA element without relying on paraxial approximations, as performed in \cite{droulias2025orthogonal}. The validation scenario considers a $1000 \times 1000$-elements array operating at $f=$ 150 GHz, with $w_0=$ 0.2 m and \mbox{$d_0$ = 10 m}. The spatial distribution of the normalized power density across the $xz$-plane is depicted in Fig. \ref{fig:spatial_distribution}, derived from the field magnitude in (\ref{eq:power_density}). Fig. \ref{fig:cross-section} illustrates a cross-section of this beam along its center ($x=y=0$), alongside with its numerically calculated counterpart. The comparison confirms that the analytical model accurately captures the longitudinal beam profile $S(z)$, correctly predicting the peak location at $z_{\text{peak}}=$ 9.75 m, and the focal extent at $w_{\text{FWHM}} =$ 3.10 m. 

The radial boundaries of the focal region are analytically derived from the $z$- axis coordinates at which peak power density $S_{\max}$ decays to a fractional threshold $\rho\in(0,1]$ of its peak value. For $N$ such beams, the parameter $\rho$ serves as a measure of the inter-beam overlap in the radial domain. Solving this condition $S(z)=\rho S_{\max}(z)$, specifically for the $n$-th focal region, where $n = 1,2,\dots N$, we can find the spatial interval of this overlap
\begin{equation}\label{eq:interval}
    \bigl[z_{\text{start},n},\; z_{\text{end},n}\bigr] = \left[ \frac{\frac{z_R^2}{d_{0,n}} - P \, z_R}{1+(\frac{z_R}{d_{0,n}})^2} ,\; \frac{\frac{z_R^2}{d_{0,n}} + P \, z_R}{1+(\frac{z_R}{d_{0,n}})^2} \right],
\end{equation}
where the parameter $P$ is defined as $P = \sqrt{\frac{1-\rho}{\rho}}$. Consequently, the difference between these two boundaries, \mbox{$\Delta z_{\rho,n}= z_{\text{end},n} - z_{\text{start},n}$}, represents the radial extent over which $S(z)\geq\rho S_{\max}$ holds for the $n$-th focal region
\begin{equation}\label{eq:delta_z}
    \Delta z_{\rho,n}= \frac{2 \,z_R}{1 + \left(\frac{z_R}{d_{0,n}}\right)^2} \, P.
\end{equation}
Evaluating at $\rho = 0.5$ yields the expression for the FWHM, consistent with the definition in \cite{droulias2024near}
\begin{equation}\label{eq:wFWHM}
    w_{\text{FWHM},n} = \frac{2 \,z_R}{1 + \left(\frac{z_R}{d_{0,n}}\right)^2}.
\end{equation}
By substituting (\ref{eq:wFWHM}) into (\ref{eq:delta_z}), we see that the radial extent $\Delta z_\rho$ for any $\rho$ scales with the beam's $w_{\text{FWHM}}$ as
\begin{equation}\label{eq:Delta_Z_of_wFWHM}
    \Delta z_{\rho, n} = P \ w_{\text{FWHM},n}.
\end{equation}

\section{Synthesis of multiple focal regions with non-uniform radial resolution using fixed beam radius}
\label{sec:same_w0}
To synthesize $N$ consecutive focal regions centered at $d_{0,1},d_{0,2},\dots,d_{0,N}$ ($d_{0,1}$ is the focal distance closest to the UPA, see Fig. \ref{fig:system_model}), we arrange the beams contiguously such that the terminal boundary of the $(n-1)$-th beam coincides exactly with the initial boundary of the $n$-th beam, as follows
\begin{equation} \label{eq:chain}
    z_{\text{end},n-1} = z_{\text{start},n} \qquad n = 2,\dots,N.
\end{equation} 

Given the focal distance $d_{0,n}$, our objective is to derive $d_{0,n-1}$ for the adjacent focal region. To enforce spatial contiguity, we combine condition (\ref{eq:chain}) with the terminal boundary expression from (\ref{eq:interval}) and rearrange the result into a quadratic equation with respect to $d_{0,n-1}$
\begin{equation} \label{eq:d_0_N-1_quadratic_eq}
    (G_n - P\,z_R)d_{0,n-1}^2 - z_R^2d_{0,n-1} + G_nz_R^2 = 0,
\end{equation} 
where $G_n = z_{\text{start},n}$ given by (\ref{eq:interval}). Solving for $d_{0,n-1}$ provides two roots, from which we select the following (see details in Appendix A)
\begin{equation} \label{eq:d_0_N-1_closed_expression}
    d_{0,n-1} = \frac{z_R \left(\sqrt{z_R^2 + 4 \,G_n \,P \,z_R - 4 \,G_n^2} - z_R\right)}{2(P\, z_R - G_n)}.
\end{equation}
This expression can be then applied iteratively to synthesize the focal distances for all $N$ consecutive regions along the radial domain.

Although the Rayleigh range $z_R$ conventionally defines the upper bound for near-field operation, the total number of synthesized focal regions is maximized by positioning the outermost beam at an optimal $d_{0,N}$ that maximizes its initial boundary $z_{\text{start},N}$. Differentiating the initial boundary given in (\ref{eq:interval}) with respect to $d_{0,N}$ and solving for the positive root yields this optimal distance
\begin{equation} \label{eq:furthest_beam}
    d_{0,N} = z_R \left( \sqrt{P^2 +1} - P \right). 
\end{equation} 

Furthermore, the minimum focal distance $d_{0,min}$ also imposes a fundamental constraint. Because the analytical framework relies on the paraxial approximation, the minimum focal distance at which this approximation accurately represents the exact propagated field must be identified. To maintain analytical validity, we define the maximum transverse spatial frequency components ($k_x$ or $k_y$) for which the approximation error $\Delta_{k_z} = \left\vert{} k_z^{\text{exact}} - k_z^{\text{approx}} \right\vert{}$ remains negligible, where $k_z^{\text{exact}}$ is given by (\ref{eq:k_z}) and $k_z^{\text{approx}}$ by (\ref{eq:paraxial}). Mapping these spatial frequencies to the propagation angle $\theta$ via the trigonometric relation $\sin\theta = k_x/k$, and restricting the beam to small angles ($\sin\theta \approx \theta \leq \theta^{\max}$), the maximum allowable transverse spatial frequency is defined as $k_x^{\max} \approx \theta^{\max} k$.

Applying the Fourier transform defined in (\ref{eq:F1_UPA}), we analytically derive the solution for $\vert{}\hat{E}^{\text{UPA}}(k_x, k_y)\vert{}$
\begin{equation} \label{eq:mag_UPA}
    \vert{}\hat{E}^{\text{UPA}}(k_x,k_y)\vert{} = \vert{}E_0\vert{} \frac{w_0}{2\pi\hat{w}_0} \exp\left[ -\frac{k_x^2+k_y^2}{\hat{w}_0^2} \right],
\end{equation}
where $\hat{w}_0=\frac{2}{w_0}\sqrt{1+(\frac{z_R}{d_0})^2}$ is the spectral width of the beam. For the paraxial approximation to hold true, the primary spectral $k$-content must be confined within the threshold, such that $\hat{w}_0 \leq k_x^{\max}$. Evaluating this inequality yields the closed-form expression for $d_{0,min}$
\begin{equation} \label{eq:d0_min}
\begin{aligned}
     d_{0,\text{min}}  &\approx \frac{2z_R}{\sqrt{2\,(\theta^{\text{max}})^2\,k\,z_R  - 4}}.
\end{aligned}
\end{equation}

A reliable boundary for $\theta^{\max}$ is established by quantifying its influence through a comparison of the exact numerical model against the analytical expressions. Specifically, we evaluate the relative error of the focal region's $w_{\text{FWHM}}$ at the corresponding $d_{0,\min}$, where the paraxial approximation error is maximized.
\begin{equation}
    \text{Relative Error}_{\%} = \left\vert{} \frac{w_{\text{FWHM}}^{\text{exact}} - w_{\text{FWHM}}^{\text{approx}}}{w_{\text{FWHM}}^{\text{exact}}} \right\vert{} \times 100.
\end{equation}

\begin{figure}
    \centering    \includegraphics[width=1.0\linewidth]{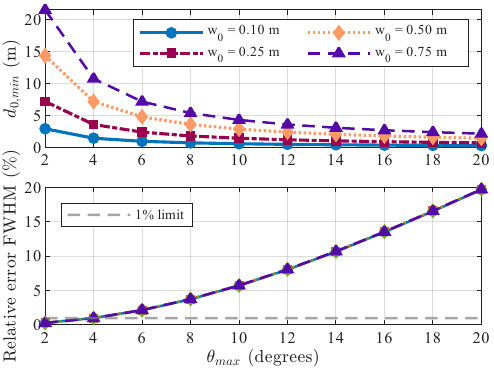}
    \caption{Analysis of the minimum focal distance and the relative error under varying maximum angular deviations. The top panel illustrates the minimum focal distance as a function of the angle across different beam radii. The bottom panel depicts the relative error of the radial extent $w_{\text{FWHM}}$ at $d_{0,\text{min}}$, highlighting the $1\%$ error limit.}
    \label{fig:theta_max}
\end{figure}

The impact of $\theta^{\max}$ on $d_{0,\min}$ across varying beam radii $w_0$ is depicted in Fig \ref{fig:theta_max}. Restricting $\theta^{\max}$ for small angles enforces greater minimum focal distance, which scales as the beam radius increases. For each computed $d_{0,\min}$ in this domain, the exact and approximated $w_{\text{FWHM}}$ are compared to measure the relative error. The approximation error remains identical regardless of the chosen design parameter $w_0$. Across all evaluated scenarios, the relative error reaches the 1\% limit at $\theta^{\max} = 4^\circ$, establishing this angle as a reliable design rule for the analytical model. Further details validating this bound are detailed in Appendix \ref{sec:appendix B}. 

To illustrate the proposed framework, we evaluate a case study comprising a square UPA transmitter with $L_x$~=~$L_y$=~1500 elements, inter-element spacing of $\frac{\lambda}{2}$, operating frequency $f=$ 150 GHz, and beam radius $w_0=$ 0.2 m. The objective is to synthesize $N$ focal regions that intersect precisely at their FWHM ($\rho =$ 0.5).

\begin{figure}[t!]
    \centering
    \includegraphics[width=\columnwidth]{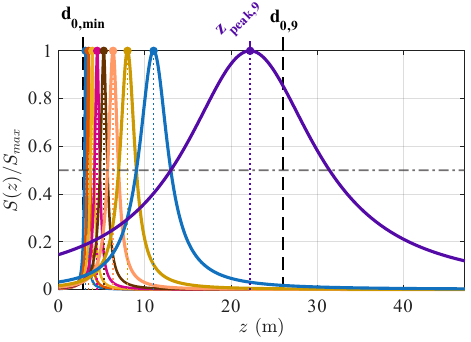}
    \caption{Resolving the radial domain with $N=9$ focal regions with non-uniform resolution. The spatial distribution illustrates contiguity at $\rho = 0.5$ inter-beam overlap under a fixed beam radius configuration.}
    \label{fig:study_case_1}
\end{figure}

As depicted in Fig. \ref{fig:study_case_1}, we distribute $N=$ 9 contiguous focal regions bounded by the optimal focal distance $d_{0,N} \approx$ 26.02 m and the minimum focal distance $d_{0,min} \approx$ 2.87 m. The response highlights the focal shift phenomenon inherent to near-field beamfocusing: although the farthest beam is targeted at $d_{0,N}$, its maximum power density shifts inward to approximately 22.21 m. Governed by (\ref{eq:z_peak}), this displacement toward the transmitter becomes more pronounced at larger focal distances (higher $d_0/z_R$ ratios). Furthermore, because $w_0$ remains fixed, the radial resolution is non-uniform; focal regions synthesized further from the UPA exhibit a broader $w_{\text{FWHM}}$ compared to those near the aperture. 

\section{Synthesis of multiple focal regions with uniform radial resolution using tunable beam radius}
\label{sec:different_w0}
\begin{figure*}[!t]
    \centering    
    \subfloat[]{        \includegraphics[width=0.45\linewidth]{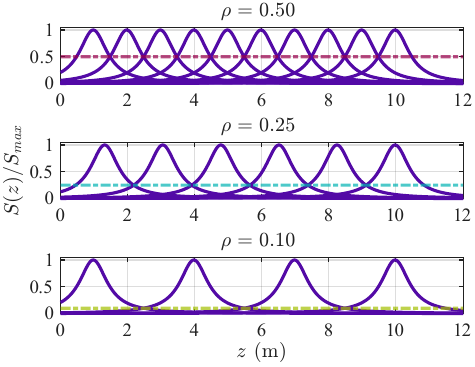}
        \label{fig:same_areas}
    }
    \hfil 
    \subfloat[]{
     \includegraphics[width=0.45\linewidth]{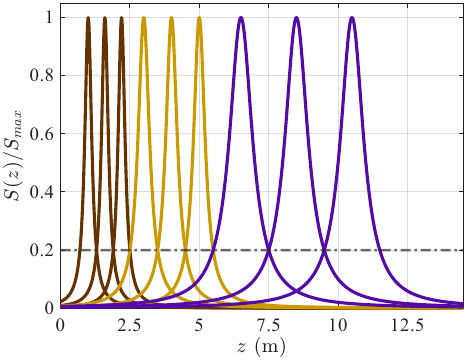}
        \label{fig:special_same}
    }    
    \caption{Resolving the radial domain with $N$ uniform focal regions. (a) Fully uniform resolution as a function of the inter-beam overlap $\rho$. (b) Block-uniform resolution for $\rho=$ 0.2.}
    \label{fig:special_cases_combined}
\end{figure*}

A uniform radial resolution is desirable when resolving the radial domain for advanced near-field applications \cite{droulias2024near, stratidakis2025near, banerjee2026volumetric}. When a constant radial extent $w_{\text{FWHM}}$ (i.e., $\Delta z_{\rho = 0.5}$) is required across all varying focal distances, the beam radius $w_0$ must be adjusted independently for each individual beam. However, for practical analytical formulation, we control $w_0$ through its associated Rayleigh range       $z_R = \frac{k w_0^2}{2}$. 

Assuming a known operation frequency $f$ and fixing the focal distance $d_{0,N}$ for the outermost region, we seek to maintain a sequence of $N$ focal regions with a constant $w_{\text{FWHM}}$ that intersect at the inter-beam overlap $\rho$. The desired radial extent $w_{\text{FWHM}}$ maps to the spatial interval $\Delta z_\rho$ via (\ref{eq:Delta_Z_of_wFWHM}). To initialize this process, we determine $z_{R,N}$ corresponding to the $N$-th focal region by solving (\ref{eq:delta_z}) for $z_R$. This establishes its initial boundary $G_N = z_{\text{start},N}$, providing the starting condition required to proceed with the recursive framework for calculating the remaining focal distances    $d_{0,n-1}$ and Rayleigh ranges $z_{R,n-1}$ for every region in the sequence. Consequently, the following system of equations is defined using (\ref{eq:chain}) and (\ref{eq:delta_z})
\begin{equation} \label{eq:system_of_equations}
    \begin{cases}
    \frac{z_{R,n-1}^2 d_{0,n-1} + P z_{R,n-1} d_{0,n-1}^2}{z_{R,n-1}^2 + d_{0,n-1}^2} = G_n, \\[8pt]
    \frac{2 z_{R,n-1} d_{0,n-1}^2 P}{z_{R,n-1}^2 + d_{0,n-1}^2} = \Delta z_\rho.
    \end{cases}
\end{equation}
Solving this system yields
\begin{align} 
    d_{0,n-1} &= \frac{\left(G_n - \frac{\Delta z_\rho}{2}\right)^2 + \left(\frac{\Delta z_\rho}{2P}\right)^2}{G_n - \frac{\Delta z_\rho}{2}} ,\label{eq:d0_n-1} \\
    z_{R,n-1} &= \frac{\left(G_n - \frac{\Delta z_\rho}{2}\right)^2 + \left(\frac{\Delta z_\rho}{2P}\right)^2}{\frac{\Delta z_\rho}{2P}}. \label{eq:zR_n-1}
\end{align}
The beam radius at the UPA for synthesizing the $(n-1)$-th focal region is given by
\begin{equation} \label{eq:w0_N-1}
w_{0,n-1} = \sqrt{\frac{2 z_{R,n-1}}{k}}.
\end{equation}

Furthermore, dynamically tailoring $z_{R,N}$ at large focal distances $d_{0,N}$ causes the beam radius $w_{0,N}$ to grow arbitrarily large. Thus, the finite UPA aperture becomes a primary operational constraint. To ensure that the Gaussian amplitude profile $A(x,y)$ remains valid, the radiated field must avoid truncation by decaying to a negligible amplitude at the array edges \cite{droulias2022reconfigurable}. This condition can be quantified by the edge taper $T_e$ \cite{goldsmith1987radiation, renker2012antenna}, evaluated boundary midpoints $(\pm\frac{l}{2}, 0)$ and $(0, \pm\frac{l}{2})$ of a square UPA of length $l$, where the edge field intensity is maximized
\begin{align} 
    \label{t_e_max}
    T^{\max}_{e} &= \frac{\vert A\left(\frac{l}{2},0\right) \vert^2}{ \vert A(0,0) \vert^2} = e^{-2\left(\frac{l}{2w_0} \right)^2}.
\end{align}
To prevent significant truncation, the array length is commonly restricted to  $l = 4w_0$ \cite{renker2012antenna}, yielding a -34.7 dB edge taper with the boundary intensity dropping to 0.034\% of the peak. Consequently, the maximum achievable beam radius is constrained to $w_{0,\max} = l/4$, which directly bounds the outermost focal region. By isolating $d_{0,N}$ in (\ref{eq:delta_z}) and substituting the corresponding maximum Rayleigh range $z_{R,\max}$, we derive the farthest aperture-constrained focal distance (examples of this boundary condition are provided in Appendix \ref{sec:appendix C})
\begin{equation} \label{eq:furthest_beam_algo_2}
    d_{0,N} = \sqrt{\frac{z_{R,\max}^2 \ \Delta z_\rho }{2 P z_{R,\max} - \Delta z_\rho }}.
\end{equation}
Furthermore, the minimum focal distance constraint $d_{0,\text{min}}$ must still be considered. Although it is evaluated using the same formulation from Section \ref{sec:same_w0}, this lower bound is dynamically updated for each focal region based on its specific Rayleigh range $z_{R,n}$.

Building upon these derived constraints, we first analyze a configuration that maintains a uniform radial extent $w_{\text{FWHM}}$ across all $N$ focal regions by tailoring the individual beam radius $w_0$. For $w_{\text{FWHM}}=$ 1 m at $d_{0,N} =$ 10 m, the Rayleigh range $z_{R,N}$ equals to 199.5 m, which corresponds to a beam radius of 0.35 m. Fig. \ref{fig:same_areas} illustrates how varying inter-beam overlap $\rho$ impacts the system multi-access capacity. At~$\rho=$~0.5, the radial interval $\Delta z_{0.5}$ coincides with the FWHM, accommodating 10 focal regions. A stricter $\rho$ demands wider spatial intervals to reach the lower intersection thresholds. Setting $\rho=$ 0.25 expands this interval to $\Delta z_{0.25}~\approx$~1.73~m, and reduces the sequence to 6 regions. A further overlap reduction to 10\% expands $\Delta z_{0.1}$ to 3 m, limiting $N$ to 4 focal regions. 

A more complex scenario evaluates a variable $w_{\text{FWHM}}$ across distinct blocks of focal regions. To support heterogeneous requirements, a generalization of (\ref{eq:d0_n-1}) and (\ref{eq:zR_n-1}) substitutes $\Delta z_\rho$ with a block-specific specific interval $\Delta_n z_\rho$. We define an arrangement of 9 focal regions grouped by their target radial extent: 0.3~m for the three innermost zones, 0.5 m for the middle three, and 1 m for the outermost three. A $\rho$ of 0.2 maintains spatial contiguity across the entire sequence.
Synthesis begins by constraining the beam radius of the farthest region to the array's physical limit ($w_{0,N} = w_{0,\max}$), which yields $z_{R,N} = z_{R,\max}=$ 220.9 m. Applying (\ref{eq:furthest_beam_algo_2}) with a target $w_{\text{FWHM},N}$ of 1~m establishes $d_{0,N} =$~10.52~m. As Fig. \ref{fig:special_same} demonstrates, all focal regions achieve their specified radial resolutions and intersect at the desired $\rho$ overlap. 

\section{Analysis of radial resolution capabilities}
\label{sec:system_analysis}
\begin{figure*}[t!] 
    \centering
    
    \subfloat[]{%
        \includegraphics[width=0.32\linewidth]{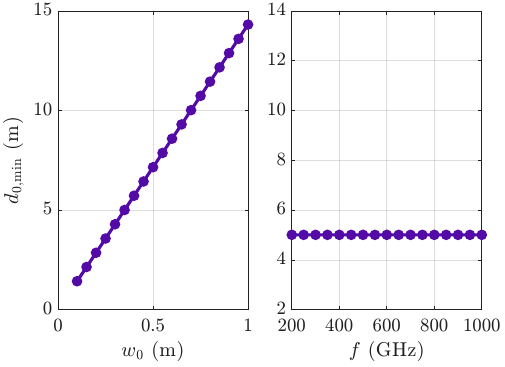}
        \label{fig:analysis_dmin_2}
    }
    \hfill 
    \subfloat[]{%
       \includegraphics[width=0.32\linewidth]{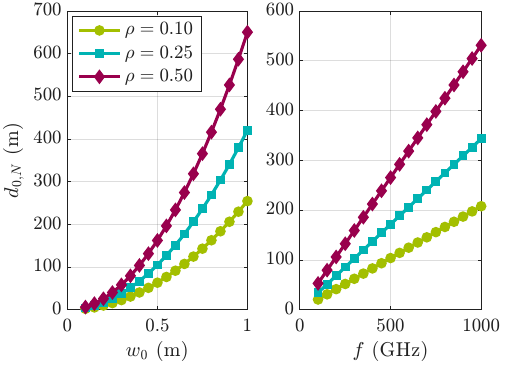}
        \label{fig:analysis_dn}
    }
    \hfill 
    \subfloat[]{%
        \includegraphics[width=0.32\linewidth]{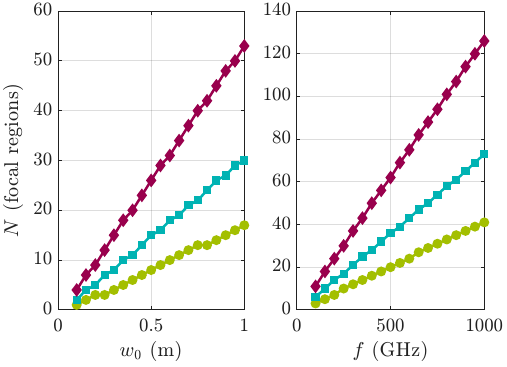}
        \label{fig:analysis_N}
    }
    
    \caption{Resolution capabilities of fixed beam radius (non-uniform radial resolution) for different input radii and operating frequencies. (a) Minimum focal distance. (b) Outermost focal distance across different $\rho$. (c) Total number of focal areas across different $\rho$. For evaluations where the beam radius varies, the operating frequency is fixed at 150 GHz; conversely, when the operating frequency varies, the beam radius is maintained at 0.2 m.}    
    \label{fig:analysis_parametric_al1}
\end{figure*}

\begin{figure}[t!]
    \centering
    \subfloat[]{%
        \includegraphics[width=0.48\columnwidth]{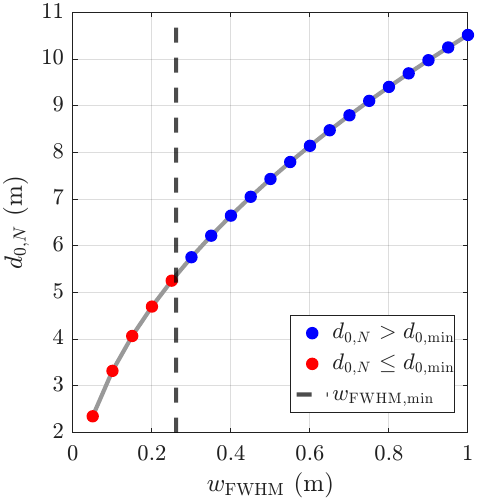}%
        \label{fig:d0_N_wFWHM}%
    }\hfill
    \subfloat[]{%
        \includegraphics[width=0.48\columnwidth]{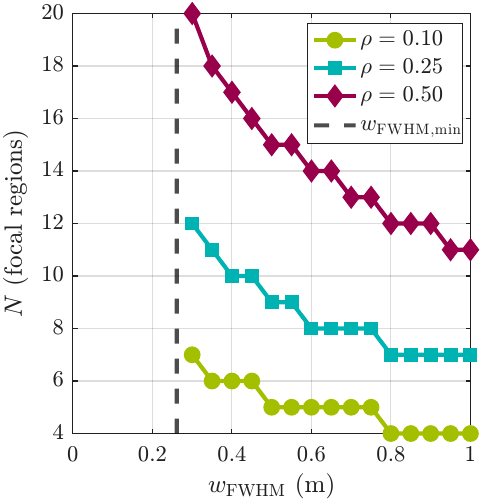}%
        \label{fig:N_wFWHM}%
    }
    \caption{Resolution capabilities of enforcing a uniform radial resolution ($w_{\text{FWHM}}$) by tailoring the beam radius on the outermost focal distance $d_{0,N}$ and the total number of focal areas $N$.}
    \label{fig:influence_wFWHM}
\end{figure}

The near-field resolution capabilities of the proposed framework are determined by the interplay of its primary design parameters. Building upon the configuration described in Section \ref{sec:same_w0}, Fig. \ref{fig:analysis_parametric_al1} illustrates how the beam radius $w_0$ and the operational frequency $f$ dictate the minimum focal distance $d_{0,\min}$, the outermost focal distance $d_{0,N}$, and the total number of synthesized focal regions $N$. To evaluate these interactions, the system is analyzed across different inter-beam overlaps $\rho \in \{0.10, 0.25, 0.50\}$, maintaining a maximum angular deviation of $\theta^{\max} = 4^{\circ}$.

The behavior of $d_{0,\min}$ is governed by the array's beam parameter $w_0$ rather than the operating frequency $f$. This relation can be analytically understood through the denominator in (\ref{eq:d0_min}). For $2(\theta^{\max})^2 k z_R \gg 4$ and substituting the definition of $z_R$, the minimum focal distance approximately simplifies to $w_0 / \theta^{\max}$. This simplification establishes a linear dependence on $w_0$ and complete independence from $f$. As verified in Fig. \ref{fig:analysis_dmin_2}, $d_{0,\min}$ scales linearly from 1.43~m to 14.32~m when $w_0$ increases from 0.1~m to 1.0~m; moreover, it also demonstrates that $d_{0,\min}$ remains constant at 5~m across the entire frequency sweep. 

In contrast, the outermost focal distance $d_{0,N}$, established in (\ref{eq:furthest_beam}), is proportional to the Rayleigh range $z_R$. Because $d_{0,N}$ scales quadratically with $w_0$ and linearly with $f$, it remains sensitive to these parameters alongside the inter-beam overlap $\rho$. Fig. \ref{fig:analysis_dn} shows that larger $w_0$ combined with $\rho$ of 0.5 positions $d_{0,N}$ to approximately 650 m, compared to just 255 under stricter inter-beam overlap. A comparable analysis across the frequency range confirms that $d_{0,N}$ displays a linear expansion on $f$, with the rate of this scaling dictated by $\rho$.

Furthermore, the total focal regions $N$ governed by the interaction between $d_{0,\min}$ and $d_{0,N}$ depicts a positive growth trend with both $w_0$ and $f$. For instance, setting $w_0=~$~1.0~m supports 53 focal regions at $\rho = 0.50$, while isolation levels of $\rho=$ 0.25 and $\rho=$ 0.10 m reduce this capacity to 30 and 17 regions, respectively. Similarly, scaling up the operating frequency expands the available radial space for user allocation. At 1000 GHz, the system accommodates up to 126 focal regions when $\rho =$ 0.50, 73 when $\rho =$ 0.25, and 41 when $\rho =$ 0.10.

Assessing the system within the variable beam radius $w_0$ and uniform-resolution framework from Section \ref{sec:different_w0} also provides insights into the interaction between $w_{\text{FWHM}}$, $d_{0,N}$, and $N$. When the outermost focal distance $d_{0,N}$ is constrained to the maximum beam radius $w_{0,\max}$, it causes $d_{0,N}$ to scale monotonically with $w_{\text{FWHM}}$. Furthermore, because the propagation axis is lower-bounded by $d_{0,\min}$, substituting (\ref{eq:d0_min}) into (\ref{eq:wFWHM}) yields the highest achievable radial resolution (i.e., the minimum focal extent $w_{\text{FWHM},\min}$) that satisfies the paraxial constraint
\begin{equation} \label{eq:w_FWHM_min}
w_{\text{FWHM},\min} = \frac{2 z_R}{1+\left(\frac{z_R}{d_{0,\min}}\right)^2} = \frac{4}{k (\theta^{\max})^2} = \frac{2 \lambda}{\pi (\theta^{\max})^2},
\end{equation}
which is dictated by the wavelength $\lambda$ and the maximum approximation error $\theta^{\max}$. For example, when $f=$ 150 GHz and $\theta^{\max} = 4^\circ$, $w_{\text{FWHM},\min}$ approximates to 0.26 m. Synthesizing focal regions narrower than this bound forces $d_{0,N} < d_{0,\min}$, which falls into the invalid operational regime visualized in  Fig. \ref{fig:d0_N_wFWHM}.
Within the valid range $w_{\text{FWHM}} > w_{\text{FWHM},\min}$, the number of focal regions $N$ exhibits an inverse relationship with $w_{\text{FWHM}}$. Near the lower bound (e.g. $w_{\text{FWHM}} =$~0.3 m), $N$ is maximized at 20, 12 and 7 supportable focal regions for inter-beam overlap of 0.50, 0.25, and 0.10, respectively.
These dynamics underscore a relevant trade-off: although increasing $w_{\text{FWHM}}$ pushes $d_{0,N}$ further away and expands the available radial space, a broader $w_{\text{FWHM}}$ simultaneously widens the spatial interval $\Delta z_\rho$ required for each individual region. Consequently, the system's capacity to synthesize contiguous focal areas decreases, as demonstrated by the $N$ behavior in Fig. \ref{fig:N_wFWHM}.

\section{System-level performance assessment}
\label{sec:system_evaluation}
\begin{figure*}[t]
    \centering
    \subfloat[]{%
        \includegraphics[width=0.48\textwidth]{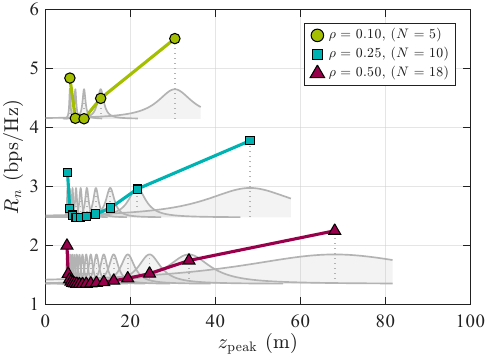}%
        \label{fig:perf_diff}%
    }%
    \hfill
    \subfloat[]{%
        \includegraphics[width=0.48\textwidth]{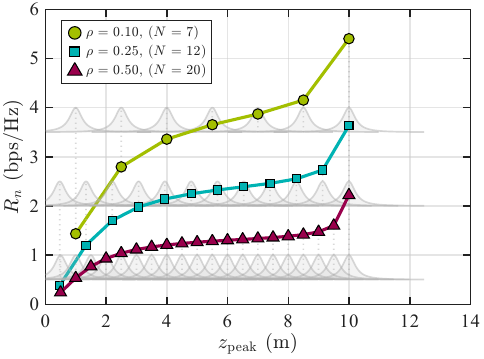}%
        \label{fig:perf_same}%
    }%
    \caption{Per-user spectral efficiency ($R_n$) at the maximum intensity point ($z_{\text{peak}}$) of each synthesized focal region under varying inter-beam overlap thresholds $\rho$. (a) Fixed beam radius (non-uniform radial resolution). (b) Tailored beam radius to maintain a uniform radial resolution.}
    \label{fig:per_user_performance}
\end{figure*}

\begin{figure*}[t]
    \centering
    \subfloat[]{%
        \includegraphics[width=0.48\textwidth]{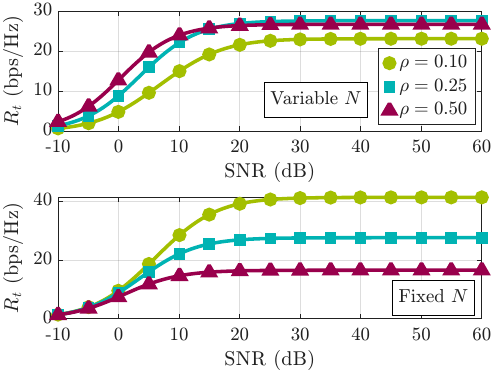}%
        \label{fig:sum_rate_fixedw0}%
    }%
    \hfill
    \subfloat[]{%
        \includegraphics[width=0.48\textwidth]{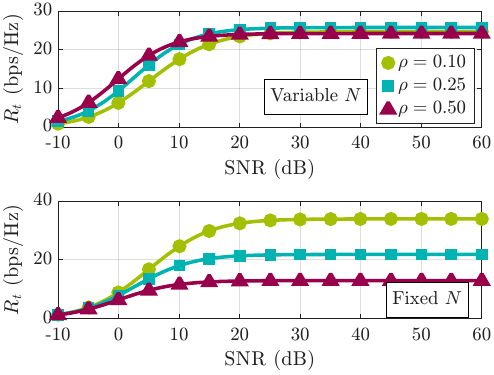}%
        \label{fig:sum_rate_differentw0}%
    }%
    \caption{System-level performance comparison evaluating the total sum rate ($R_t$) as a function of the SNR across varying inter-beam overlap $\rho$. (a) Total capacity using a fixed beam radius (non-uniform resolution). (b) Total capacity maintaining a uniform resolution $w_{\text{FWHM}}$. In both (a) and (b), the top panels depict the variable user scenario, whereas the bottom panels depict the fixed user scenario.}
    \label{fig:overall_performance}
\end{figure*}

\begin{figure}
    \centering
    \subfloat[]{%
        \includegraphics[width=1\linewidth]{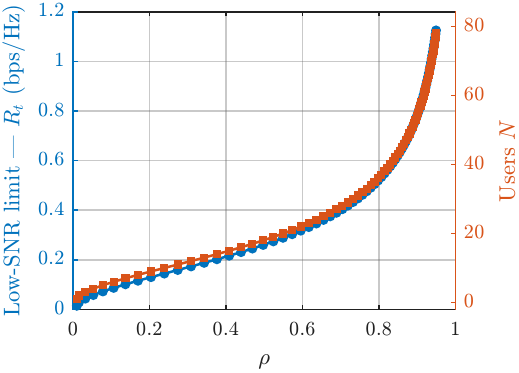}%
        \label{fig:low_SNR_regime}
    }
    
    \subfloat[]{%
        \includegraphics[width=1\linewidth]{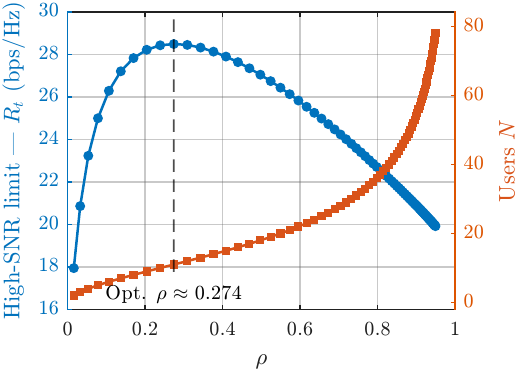}%
        \label{fig:high_SNR_regime}%
    }
    \caption{Asymptotical analysis of $R_{t}$ and user capacity $N$ as a function of the inter-beam overlap $\rho$. (a) The noise limited regime (low-SNR limit). (b) The interference-limited regime     (high-SNR limit).}
    \label{fig:overall-asym}
\end{figure} 

To assess the multiple-access capabilities of the $N$ synthesized focal regions, we evaluate the system performance in terms of achievable spectral efficiency. This assessment relies on the signal-to-interference-plus-noise ratio (SINR) computed at the peak intensity point $z_{\text{peak}}$ of each beam. Formulated as a function of the SNR, the SINR is given by
\begin{align}\label{eq:SINR}
    \text{SINR}_n &= \frac{P_n}{\sum_{l\not=n}P_l + P_{\nu}} = \ \frac{\text{SNR}}{\alpha_n \cdot \text{SNR} + 1},
\end{align}
where $P_n$ denotes the received power at the peak of the $n$-th beam, and $P_\nu$ represents the noise power, such that $\text{SNR}~=~\frac{P_n}{P_\nu}$. The parameter              $\alpha_n = \frac{\sum_{l\not=n}P_l}{P_n}$ is proportional to the level of interference experienced by the respective beam. From this SINR definition, the achievable rate $R_n$ for the $n$-th beam is calculated as
\begin{equation}\label{eq:R_m}
    R_n = \log_2(1 + \text{SINR}_n).
\end{equation}
Consequently, the sum-rate of the system $R_{\text{t}}$ is determined by aggregating the achievable rates of all individual beams
\begin{equation} \label{eq:R_total}
    R_{\text{t}} = \sum_n^N R_n.
\end{equation}

To analyze the impact of the parameter $\rho$ on the system's user capacity, we evaluate the two synthesis configurations detailed in Sections \ref{sec:same_w0} and \ref{sec:different_w0}. Both frameworks are compared under variable and fixed user-count ($N$) scenarios. Table \ref{tab:rho_cases} summarize the primary parameters.

For the variable user scenario, the per-user performance is illustrated in Fig. \ref{fig:per_user_performance}. When $w_0$ is fixed,  central focal regions exhibit the lowest spectral efficiencies across all inter-beam overlaps due to the interference from adjacent beams, as depicted in Fig. \ref{fig:perf_diff}. The innermost $d_{0,\min}$ and outermost $d_{0,N}$ distances experience less interference, thereby achieving highest transmission rates. Results reveal that whereas a larger $\rho$ accommodates more users $N$, it compromises the achievable rate $R_n$. In contrast, Fig.~\ref{fig:perf_same} shows that enforcing $w_{\text{FWHM}}=$~0.5 m by tailoring $w_0$, increases the per-user rate with focal distance. Maintaining this uniform extent further from the array requires a larger $w_0$; consequently, the $w_0^4$ peak power density scaling (via $z_R$) in (\ref{eq:s_z,max}) overcomes the inverse-square distance attenuation, which yields higher spectral efficiencies for distant regions. Nevertheless, $\rho$ still governs the trade-off between user capacity and individual rate.

To evaluate the system-level performance, we analyzed the sum-rate $R_{\text{t}}$ as a function of the SNR. For the variable user scenario, both top panels in Fig. \ref{fig:overall_performance} illustrate a similar performance trend across the syntheses configurations. In the low-SNR regime, the highest total rate is achieved by maximizing the inter-beam overlap. As the system transitions to higher SNR values, the configuration with $\rho = 0.50$ becomes severely interference-limited and saturates early; therefore, $\rho = 0.25$ overtakes it, achieving the highest overall capacity. This occurs because it provides a better balance between users supported ($N$ ranging from 10-12) and inter-user interference suppression. Furthermore, the constant user-count depicted in the bottom panels ($N$ ~= 10), indicates that the sum-rate is influenced by minimizing the inter-user interference to enhance the individual rate $R_n$. Consequently, the inter-beam overlap $\rho=$ 0.10 yields the highest performance across the SNR domain. 
\begin{table}[!t]
\caption{Summary of system performance parameters for both synthesis frameworks}
\label{tab:rho_cases}
\centering

\setlength{\tabcolsep}{3.5pt}

\textbf{(a) Case 1: Non-uniform radial resolution (fixed $w_0$)} \\*[0.5em]
\begin{tabular}{c c c c c c c}
\hline
$\rho$ & $w_0$ & $d_{0,\text{N}}$ & $d_{0,\min}$ & $N$ & $R_t$ (10 dB) & $R_t$ (60 dB) \\ 
 & \textbf{(m)} & \textbf{(m)} & \textbf{(m)} & & \textbf{(bps/Hz)} & \textbf{(bps/Hz)} \\
\hline
    $0.50$ & 0.35 & 79.70 & 5.04 & 18 & 23.95 & 26.70 \\
    $0.25$ & 0.35 & 51.56 & 5.17 & 10 & 22.07 & 27.62\\
    $0.10$ & 0.35 & 31.23 & 5.73 & 5  & 14.94 & 23.10 \\
\hline
\noalign{\vspace{0.5em}}
\hline
    $0.50$ & 0.21 & 28.69 & 3.14 & 10 & 14.60 & 16.52 \\
    $0.25$ & 0.35 & 51.56 & 5.17 & 10 & 22.07 & 27.62 \\
    $0.10$ & 0.61 & 94.85 & 9.16 & 10 & 28.48 & 41.39 \\
\hline
\end{tabular}

\vspace{1.5em}

\textbf{(b) Case 2: Uniform radial resolution (tailored $w_0$)} \\*[0.5em]
\begin{tabular}{c c c c c c c}
\hline
$\rho$ & $w_{\text{FWHM}}$ & $d_{0,\text{N}}$ & $d_{0,\min}$ & $N$ & $R_t$ (10 dB) & $R_t$ (60 dB) \\ 
 & \textbf{(m)} & \textbf{(m)} & \textbf{(m)} & & \textbf{(bps/Hz)} & \textbf{(bps/Hz)} \\
\hline
    $0.50$ & 0.5 & 10 & 1.23 & 20 & 22.04 & 24.24 \\
    $0.25$ & 0.5 & 10 & 1.12 & 12 & 21.38 & 25.78 \\
    $0.10$ & 0.5 & 10 & 1.06 & 7  & 17.54 & 24.67 \\
\hline
\noalign{\vspace{0.5em}}
\hline
    $0.50$ & 1.0 & 10 & 1.23 & 10 & 11.63 & 12.96 \\
    $0.25$ & 0.6 & 10 & 0.78 & 10 & 18.02 & 21.86 \\
    $0.10$ & 0.3 & 10 & 0.6  & 10 & 24.56 & 33.93 \\
\hline
\end{tabular}
\end{table}

\subsection{Asymptotic analysis of the sum-rate $R_{\text{t}}$ for the variable user-count scenario}
The performance dynamics of the variable user case reveal fundamental trade-offs involved in near-field multiple access applications. 
To characterize how the design parameter $\rho$ shifts the balance between supporting more users and mitigating inter-user interference, we perform an asymptotic analysis when focal areas have fixed $w_0$ (top panel Fig. \ref{fig:sum_rate_fixedw0}). However, an analogous evaluation can be applied when dynamically tailoring $w_0$ to synthesize focal areas of uniform radial extent.

In the low-SNR regime, the interference term $\alpha_n\text{SNR}$ in (\ref{eq:SINR}) becomes negligible, simplifying the denominator to unity. Consequently, the SINR approximates to $\frac{\text{SNR}}{\alpha_n(\rho) \text{SNR} + 1} \approx \text{SNR}$. Applying the first-order Maclaurin series expansion for the logarithm ($\log_2(1+x) \approx \frac{x}{\ln 2}$ for $x \to 0$), the sum-rate is derived as
\begin{align}\label{eq:R_total_low_SNR}
    R_{t} &\approx \sum_{n=1}^{N} \log_2 ( 1 + \text{SNR} ) 
    &&\approx \sum_{n=1}^{N} \frac{\text{SNR}}{\ln 2} 
    &&= N\frac{\text{SNR}}{\ln 2}.
\end{align}
The total sum-rate is noise-limited and scales linearly with $N$, independent of the interference scalar $\alpha_n$. Therefore, increasing $\rho$ to accommodate more users maximizes the total capacity. Fig. \ref{fig:low_SNR_regime} illustrates this behavior at an SNR of -20 dB, confirming that the spectral efficiency grows monotonically as more users are supported within the radial domain.

Conversely, in the high-SNR regime ($\text{SNR} \to \infty$), the SINR converges to the limited bound of                 $\lim_{\text{SNR}~\to~\infty}~\frac{1}{\alpha_n + 1/\text{SNR}} =~\frac{1}{\alpha_n}$. Substituting this limit into the sum rate expression yields the asymptotic total rate
\begin{equation}\label{eq:R_total_high_SNR}
    \lim_{\text{SNR} \to \infty} R_{t} = \sum_{n=1}^{N} \log_2 \left( 1 + \frac{1}{\alpha_n} \right).
\end{equation}
The analysis demonstrates that high-SNR scenarios are interference-limited, establishing an operational trade-off. Whereas increasing $\rho$ accommodates larger user count $N$, it simultaneously amplifies the inter-user interference $\alpha_n$, which penalizes the per-user spectral efficiency dictated by the $\log_2(1 + 1/\alpha_n)$ term. Consequently, adding users only improves the total sum rate if the inter-beam overlap gains exceed these interference penalties. This dynamic is depicted in Fig. \ref{fig:high_SNR_regime} for $f=$ 150 GHz and $w_0=$~ 0.35 m, where the total capacity exhibits a concave trajectory. Although expanding $\rho$ increases $N$, the escalating inter-user interference ultimately causes a sharp decline in the overall capacity. As a result, we find an optimal configuration at $\rho \approx$ 0.27, which balances user density and interference mitigation to achieve a maximum spectral efficiency of approximately 28.5 bps/Hz within an~11-user~system.

\section{Conclusions}
\label{sec:conclu}
This work proposed an analytical framework based on the ASR to synthesize and manage multiple focal regions along the radial domain. We formulated close-form expressions to analyze the radial resolution of these regions, accommodating configurations with either a fixed beam radius or dynamically tailored radial extents. The explicit formulation of the operational boundaries, namely the minimum focal distance $d_{0,min}$ and the outermost focal distances $d_{0,N}$, define the model system capacity $N$, which is governed by the inter-beam overlap parameter $\rho$. Evaluating these resolution capabilities reveals critical interactions between the operating frequency $f$ and beam radius $w_0$, effectively translating complex-near field propagation dynamics into direct analytic design rules. Furthermore, a performance assessment demonstrates that $\rho$ dictates a decisive multiple access trade-off: balancing user throughput maximization in noise-limited scenarios against multi-user interference mitigation in capacity-limited conditions. The practical significance of this framework is anchored in the need for fundamental metrics to accurately characterize the baseline performance of emerging technologies, thereby providing scalable design guidelines for the deployment of beamfocusing in next-generation systems.

\appendices

\section{Derivation of the positive root domain for $d_{0,N-1}$}
\label{sec:appendix A}
The quadratic equation governing the adjacent focal distance $d_{0,n-1}$ from (\ref{eq:d_0_N-1_closed_expression}) can be expressed in the standard form $ax^2 + bx + c = 0$, with the coefficients defined as $a = G_n - Pz_R$, $b = -z_R^2$, and $c = G_n z_R^2$. To determine the sign of the roots $r_1$ and $r_2$ from (\ref{eq:d_0_N-1_closed_expression}) using Vieta's formula, the sign of the leading coefficient $a$ must first be established. Substituting the definition of $G_n = z_{\text{start},n}$ from (\ref{eq:interval}) and introducing the variable $u = \frac{z_R}{d_{0,n}}$ yields
\begin{equation}
    G_n - Pz_R = \frac{z_R[-Pu^2 + u - 2P]}{1 + u^2}.
\end{equation}
The denominator $1+u^2$ and the Rayleigh range $z_R$ are positive, thus the sign depends on the quadratic numerator $f(u) = -P u^2 + u - 2P$. The discriminant of $f(u)$ is given by $\Delta_u = 1 - 8P^2$. Since $f(u)$ is a downward-opening parabola (as $-P < 0$), the condition $\Delta_u < 0$ guarantees that $f(u) < 0$ for all real $u$, which holds when $P > \frac{1}{\sqrt{8}}$. Given that $P$ is defined as $P = \sqrt{\frac{1-\rho}{\rho}}$, this condition translates to an inter-beam overlap of $\rho < 0.89$. Therefore, the coefficient $a$ is strictly negative when $\rho < 0.89$, which covers a reasonable range of practical inter-beam overlap.

With $a<0$, the sum of the roots $r_1 + r_2$ is strictly negative as $r_1 + r_2= \frac{-b}{a} = \frac{z_R^2}{G_n - P z_R} <0$. Furthermore, the product of the roots $r_1 \cdot r_2 = \frac{c}{a} = \frac{G_n z_R^2}{G_n - P z_R}$ is also negative, as $G_n > 0$ because it represents a positive distance. With both the sum and the product yielding a negative value, there is exactly one positive solution and one negative solution for $d_{0,n-1}$. Selecting the positive solution requires choosing the branch that produces a negative numerator, as the denominator $2a$ is negative. As the positive branch always yields a positive numerator $z_R \left(z_R + \sqrt{z_R^2 + 4 \,G_n \,P \,z_R - 4 \,G_n^2}\right)$ and $\sqrt{z_R^2 + 4G_n(Pz_R - G_n)} > z_R$, the negative branch provides the required positive solution for $d_{0,n-1}$.

For the region of $\rho > 0.89$ the coefficient $a>0$, which results in $r_1+r_2 > 0$ and $r_1 \cdot r_2 >0$, indicating that two positive solutions for $d_{0,n-1}$ exist. In this scenario, the positive branch is discarded as it yields $d_{0,n-1} > d_{0,n}$, meaning the negative branch consistently provides the solution for $d_{0,n-1}$ as expressed in (\ref{eq:d_0_N-1_closed_expression}).

\section{Analytical limitations imposed by $\theta^{\text{max}}$}
\label{sec:appendix B}
\begin{figure}[t]
    \centering

    \subfloat[]{%
        \includegraphics[width=0.48\linewidth]{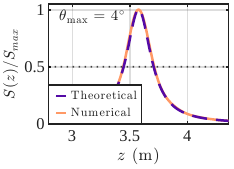}%
        \label{fig:row1_long}%
    }%
    \hfill
    \subfloat[]{%
        \includegraphics[width=0.48\linewidth]{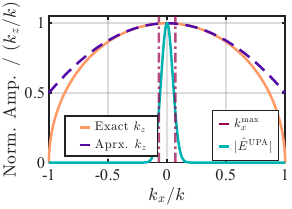}%
        \label{fig:row1_spec}%
    }

    \vspace{0.3em}

    \subfloat[]{%
        \includegraphics[width=0.48\linewidth]{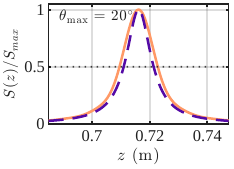}%
        \label{fig:theta_4}%
    }%
    \hfill
    \subfloat[]{%
        \includegraphics[width=0.48\linewidth]{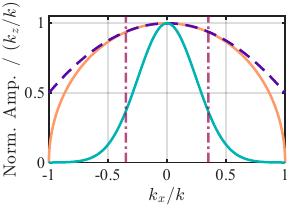}%
        \label{fig:theta_20}%
    }

    \caption{Assessment of analytical limitations dictated by the maximum angular deviation ($\theta^{\text{max}}$). The left column compares the theoretical and numerical longitudinal beam profiles, while the right column depicts the corresponding spatial spectral cuts. (a)-(b) $\theta^{\text{max}} = 4^\circ$. (c)-(d) $\theta^{\text{max}}~=~20^\circ$.}
    \label{fig:appendix_2}
\end{figure}

This appendix illustrates the influence of the maximum angular parameter $\theta^{\max}$ on the accuracy of the analytical framework. Fig. \ref{fig:appendix_2} depicts the power density and the spatial spectrum of the focal region at $d_{0,\min}$ for $\theta^{\max}$ = 4° and $\theta^{\max}$~=~20°, operating with a same $w_0$ = 0.25~m and $f=$~150~GHz. 

The results demonstrate that the error introduced by the approximated $k_z^{\text{approx}}$ remains negligible near the propagation axis ($k_x = 0$) but amplifies at higher spatial frequencies. The strongest threshold of $\theta^{\max}$ = 4°, ensures that the maximum transverse spatial frequency $k_x^{\max}$ confines the spectral components to this low-error region, maintaining their close proximity to the propagation axis for $k_z^{\text{approx}}$ to remain accurate.
Conversely, loosening this constraint to $\theta^{\max}$ = 20° produces a  broader spectral beam whose components extend in the region where the approximation's precision degrades.

The sensitivity of the model to $\theta^{\max}$ is evident on the power density plots for each distinct $d_{0,min}$. For $\theta^{\max}$ = 4°, the analytical model exhibits excellent agreement with the numerical one. However, at $\theta^{\max}$ = 20°, our framework starts to diverge indicating that the error introduced by the $k_z^{\text{approx}}$ is no longer negligible.

\section{Validation of the array boundary condition}
\label{sec:appendix C}
\begin{figure}[t!]
    \centering

    \subfloat[]{%
        \includegraphics[width=0.48\linewidth]{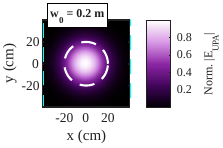}%
        \label{fig:footprint1}%
    }%
    \hfill
    \subfloat[]{%
        \includegraphics[width=0.48\linewidth]{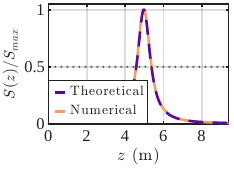}%
        \label{fig:footprint2}%
    }

    \vspace{0.3em}

    \subfloat[]{%
        \includegraphics[width=0.48\linewidth]{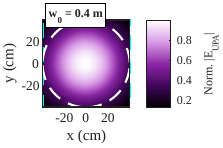}%
        \label{fig:power_density1}%
    }%
    \hfill
    \subfloat[]{%
        \includegraphics[width=0.48\linewidth]{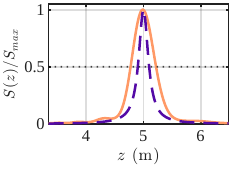}%
        \label{fig:power_density2}%
    }

    \caption{Validation of the maximum beam radius constraint $w_{0,\text{max}} = l/4$. The left column displays the normalized aperture field magnitude ($\vert{}E_{\text{UPA}}\vert{}$), whereas the right column shows the corresponding normalized power density ($S(z)/S_{\text{max}}$). (a)-(b) A beam radius of $w_0 = 0.2$ exhibits minimal boundary truncation. (c)-(d) Expanding $w_0$ to 0.4 m leads to significant aperture truncation.}
    \label{fig:appendix_3}
\end{figure}

This appendix validates the physical constraints of the system, specifically studying how the finite aperture size limits the maximum achievable beam radius $w_0$.

Fig. \ref{fig:footprint1} and Fig. \ref{fig:footprint2} depict the magnitude of the field amplitude $A(x,y)$ at the aperture plane. For a UPA comprising $800 \times 800$ elements operating at 150 GHz with half-wavelength inter-element spacing ($\lambda/2$), the resulting the aperture length is is $l = L_x \frac{\lambda}{2} \approx$ 0.8 m. Enforcing the design limit   $w_0 = w_{0,\max} = l/4 =$ 0.2 m ensures that the beam radius is fully contained within the array dimensions, as shown in Fig. \ref{fig:footprint1}. This configuration produces a strong edge taper of \mbox{$T_e =$ -34.7 dB}, which results in negligible field amplitude at the boundaries. Consequently, at a focal distance of $d_0 =$ 5 m, we observe agreement between the analytical and numerical model (Fig \ref{fig:power_density1}).

However, when $w_0$ is expanded to  0.4 m, the beam radius is barely enclosed by the aperture. This produces a much weaker edge taper of $T_e =$ -8.68 dB, leading to field truncation at the boundary at significant levels (Fig. \ref{fig:footprint2}).  As demonstrated in Fig. \ref{fig:power_density2}, this truncation causes the analytical model to diverge from the numerically computed field, because the amplitude profile $A(x,y)$ can no longer be accurately modeled as a valid Gaussian footprint.

\section*{Acknowledgments}
This work was supported by the European Union (HORIZON-MSCA-2023-DN-01), Grant Agreement 101169044 - TeraWireless.
\bibliographystyle{IEEEtran}
\bibliography{bib}

\begin{IEEEbiography}[{\includegraphics[width=1in,height=1.25in,clip,keepaspectratio]{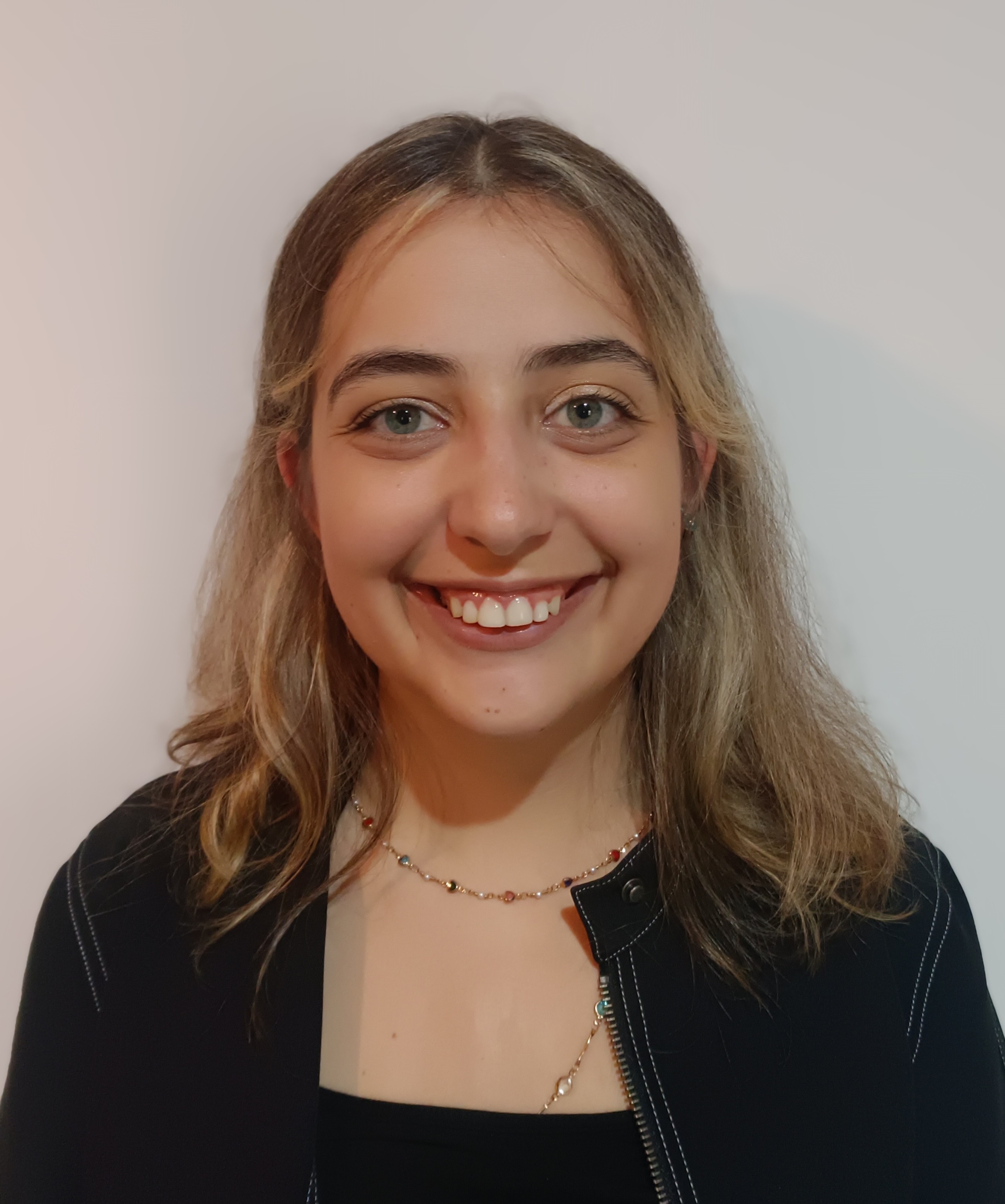}}]{VIVIANA CENTRITTO} received the B.S. degree in telecommunication engineering from Universidad Católica Andrés Bello in Caracas, Venezuela, in 2023, and the M.S. degree in advanced telecommunication technologies from Universitat Politècnica de Catalunya, Barcelona, Spain, in 2026. During her master's studies, she conducted research at the Nanonetworking Center in Catalunya (N3Cat) focusing on cryo-CMOS on-chip antenna design for quantum environments. Aditionally, she worked on the characterization of underwater wireless optical channels (UWOC) and reinforcement learning-based 3D beam adaptation for UWOC systems. She is currently a Ph.D. candidate in the Department of Digital Systems at the ICT School of the University of Piraeus, Greece, within the TeraWireless HORIZON-MSCA Industrial Doctoral Network. Her current research interests include statistical channel modeling of small scale fading channels, path resolvability, and the fundamental limit analysis of U-MIMO THz systems.
\end{IEEEbiography}

\begin{IEEEbiography}[{\includegraphics[width=1in,height=1.25in,clip,keepaspectratio]{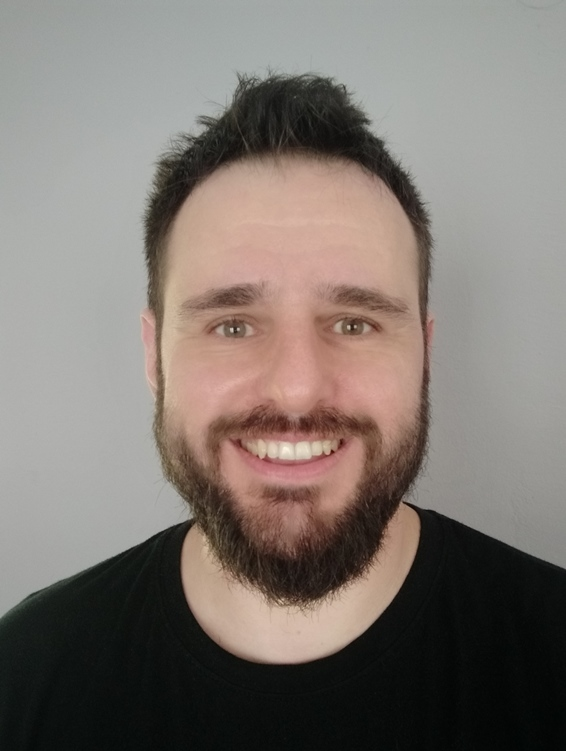}}]{SOTIRIS DROULIAS } received the diploma in electrical and computer engineering and the Ph.D. degree in nonlinear photonics from the National Technical University of Athens, Greece, in 2001 and 2007, respectively. From 2009 to 2012, he was an Adjunct Lecturer with the University of Patras, Greece, and from 2012 to 2020, he was a member with the Photonic-Phononic-and Meta-Materials Group, FORTH-IESL, Crete, Greece. He is currently a Research Associate with the Department of Digital Systems, ICT School, University of Piraeus, Greece. He is the author of more than 60 articles and four book chapters. He has worked on several EC funded projects. His research interests include electromagnetic modeling, metasurfaces, antennas, and propagation. In 2019, he received the Best Poster Award for his work on metasurface lasers in META 2019, Lisbon, Portugal, and in 2020, he was recognized as an Outstanding Reviewer from the Institute of Physics (IOP). He has received several talk invitations in prestigious conferences and he serves as a reviewer in international scientific journals.
\end{IEEEbiography}

\begin{IEEEbiography}[{\includegraphics[width=1in,height=1.25in,clip,keepaspectratio]{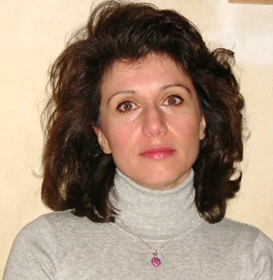}}]{ANGELIKI ALEXIOU } Angeliki Alexiou is a professor of Broadband Communications Systems at the department of Digital Systems, ICT School, University of Piraeus, Greece. She received the Diploma in Electrical and Computer Engineering from the National Technical University of Athens in 1994 and the PhD in Electrical Engineering from Imperial College of Science, Technology and Medicine, University of London in 2000. Since May 2009 she has been a faculty member at the Department of Digital Systems, where she conducts research and teaches courses on Broadband Communications and Advanced Wireless Technologies. Prior to this appointment she was with Bell Laboratories, Wireless Research, Lucent Technologies, (later Alcatel-Lucent, now NOKIA), in Swindon, UK, (January 1999-April 2009). Professor Alexiou is a co-recipient of Bell Labs President’s Gold Award in 2002 for contributions to Bell Labs Layered Space-Time (BLAST) project and the Central Bell Labs Teamwork Award in 2004 for role model teamwork and technical achievements in the IST FITNESS project. Professor Alexiou is the Chair of the Working Groups on Radio Communication Technologies and of on High Frequencies Radio Technologies of the Wireless World Research Forum and has been driving the WWRF contributions to ITU-R WP5D works on  IMT-2020  and IMT-2030. Her current research interests include radio interface for 6G systems, MIMO, THz wireless communication technologies, Reconfigurable Intelligent Surfaces, Joint Communications and Sensing, Near-Field Communications and Machine Learning for wireless systems. She is the technical manager of (SNS JU) INSTINCT project and the project coordinator of TeraWireless HORIZON-MSCA Industrial Doctoral Network.
\end{IEEEbiography}

\end{document}